\documentclass[
  journal=largetwo,
  manuscript=article-type,
  year=2020,
  volume=37,
]{cup-journal}

\usepackage{amsmath}
\usepackage[nopatch]{microtype}
\usepackage{booktabs}
\usepackage[colorlinks=true, linkcolor=blue, citecolor=blue, urlcolor=blue]{hyperref}
\usepackage{aas-macros}

\title{Growth of black holes at the centre of early-type galaxies in MOND}

\author{Robin Eappen}
\affiliation{Helmholtz-Institut für Strahlen- und Kernphysik (HISKP), Universität Bonn, Nussallee 14–16, 53115 Bonn, Germany}
\email[Robin Eappen]{robin.eappen@gmail.com}

\author{Pavel Kroupa}
\affiliation{Helmholtz-Institut für Strahlen- und Kernphysik (HISKP), Universität Bonn, Nussallee 14–16, 53115 Bonn, Germany}
\alsoaffiliation{Charles University in Prague, Faculty of Mathematics and Physics, Astronomical Institute, V Holešovickách 2, CZ-180 00 Praha 8, Czech Republic}

\bibliography{sample}
\keywords{Galaxy: evolution, Galaxy: formation, Galaxies: elliptical and lenticular, cD, Galaxies: fundamental parameters, black holes} 

\begin{document}

\begin{abstract}
The formation of supermassive black holes (SMBHs) in early-type galaxies (ETGs) is a key challenge for galaxy formation theories. Using the monolithic collapse models of ETGs formed in Milgromian Dynamics (MOND) from \citet{2022MNRAS.516.1081E}, we investigate the conditions necessary to form SMBHs in MOND and test whether these systems adhere to observed SMBH-galaxy scaling relations. We analyse the evolution of the gravitational potential and gas inflow rates in the model relics with a total stellar mass ranging from \(0.1 \times 10^{11} \, M_\odot\) to \(0.7 \times 10^{11} \, M_\odot\). The gravitational potential exhibits a rapid deepening during the initial galaxy formation phase, accompanied by high gas inflow rates. These conditions suggest efficient central gas accumulation capable of fuelling SMBH formation. We further examine the \(M_{\rm BH} - \sigma\) relation by assuming that a fraction of the central stellar mass contributes to black hole formation. Black hole masses derived from 10$\%$–100 $\%$ of the central mass are comparable with the observed relation, particularly at higher central velocity dispersions (\(\sigma > 200 \, \text{km/s}\)). This highlights the necessity of substantial inner mass collapse to produce SMBHs consistent with observations. Our results demonstrate that MOND dynamics, through the rapid evolution of the gravitational potential and sustained gas inflows, provide a favourable environment for SMBH formation in ETGs. These findings support the hypothesis that MOND can naturally account for the observed SMBH-galaxy scaling relations without invoking cold dark matter, emphasizing the importance of early gas dynamics in determining final SMBH properties.
\end{abstract}

\section{Introduction} \label{sec:intro}
Early-type galaxies (ETGs), including ellipticals and lenticulars, are compact spheroid systems characterized by old stellar populations, minimal gas content, and little to no ongoing star formation (\citealt{2009ApJS..182..216K}). These galaxies serve as vital laboratories for understanding galaxy evolution and the co-evolution of supermassive black holes (SMBHs) with their host galaxies. The scaling relations between SMBHs and galaxy properties, such as the \(M_{\rm BH}-\sigma\) (black hole mass versus central velocity dispersion) and \(M_{\rm BH}-M_{\rm bulge}\) (black hole mass versus bulge mass) relations, underscore the profound connection between SMBHs and the structural and dynamical properties of ETGs (\citealt{2013ApJ...764..184M}).

SMBHs, with masses exceeding \(10^6 \, M_\odot\), are found at the centres of most massive ETGs (\citealt{1995ARA&A..33..581K,2001RMxAC..10...69K}). Observations of high-redshift quasars reveal that these objects formed early in the universe's history, with some SMBHs already reaching masses of \(10^9 \, M_\odot\) less than a billion years after the Big-Bang (\citealt{2003AJ....125.1649F,2011Natur.474..616M,2013ApJ...779...24V}). Similarly, the stellar populations in their host galaxies also formed rapidly, with star formation timescales (SFTs) shorter than 1 Gyr (\citealt{2005ApJ...621..673T,2015MNRAS.448.3484M,2016ApJ...818..179L,2021A&A...655A..19Y}). Explaining how stellar-mass black hole seeds grow into SMBHs within these short timescales remains a major challenge. Continuous gas accretion at the Eddington limit for nearly 1 Gyr is required to achieve such growth, yet this scenario is inconsistent with the rapid cessation of gas inflows in these galaxies after star formation.

Recent studies suggest that stellar-dynamical processes may provide an alternative pathway for the formation of supermassive black holes (SMBHs). \citet{2020MNRAS.498.5652K} proposed that SMBHs can form through the merger and coalescence of stellar-mass black holes within dense star clusters. These clusters form in the central regions of collapsing gas clouds during galaxy formation, where high gas inflow rates and proto-spheroid densities drive the central cluster of stellar remnants into a relativistic regime. This process leads to catastrophic collapse through gravitational wave emission. Such a mechanism naturally accounts for the observed correlations between SMBH mass and spheroid mass.

The Milgromian Dynamics (MOND) paradigm offers a compelling framework for understanding these processes. First introduced by \citet{1983ApJ...270..365M}, MOND modifies the laws of gravity at low accelerations (\(a < a_0\), where $a_0$ $\approx$ 3.8 pc/Myr$^{2}$ is Milgrom's critical acceleration), eliminating the need for dark matter in galaxies. MOND has successfully predicted numerous galaxy-scale observations, including flat rotation curves (\citealt{1979ARA&A..17..135F, 2002ARA&A..40..263S}), the Baryonic Tully-Fisher Relation (BTFR; \citealt{2012AJ....143...40M}), and the Radial Acceleration Relation (RAR; \citealt{2017ApJ...836..152L}). Comparative studies have shown that MOND performs as well as, or better than the standard, $\Lambda$CDM model in explaining galaxy dynamics and scaling relations (\citealt{2022Symm...14.1331B}). This issue is particularly relevant given that dark matter particles have not been experimentally verified, and the dynamical friction test further questions their existence \citep{2024Univ...10..143O}. Within MOND, ETGs are thought to form via the monolithic collapse of primordial pre-galactic gas clouds, producing high stellar densities and rapid star formation (\citealt{2022MNRAS.516.1081E}). Stellar-mass black holes forming in dense central clusters during this phase may subsequently merge and coalesce to create SMBHs, as suggested by \citet{2020MNRAS.498.5652K}.

Numerical simulations in MOND using the Phantom of RAMSES (PoR) code (\citealt{2015CaJPh..93..232L,2021CaJPh..99..607N}) have reproduced many properties of observed galaxies, including the emergence of the downsizing trend in ETGs (\citealt{2022MNRAS.516.1081E}). These simulations provide a platform for studying the formation and evolution of ETGs under MOND and their potential to host SMBHs. However, while the downsizing relations and star formation histories of ETGs in MOND have been explored, the role of the central gravitational potential (\(\Phi_{\rm central}\))—a key determinant of SMBH formation—remains under explored. Observations have shown that SMBH properties strongly correlate with central galaxy dynamics, linking \(\Phi_{\rm central}\) to velocity dispersion, gas inflow, and scaling relations such as the \(M_{\rm BH}-\sigma\) relation (\citealt{2013ApJ...764..184M}).

In this study, we investigate \(\Phi_{\rm central}\) and gas inflow rates in ETGs formed in MOND simulations using the PoR code. We aim to determine whether these systems create the necessary conditions for SMBH formation and whether the resulting SMBH masses are consistent with the observed \(M_{\rm BH}-\sigma\) scaling relation. To achieve this, we analyse the evolution of the central gravitational potential and gas inflow rates during the early phases of galaxy formation. Additionally, based on the work of \citet{2020MNRAS.498.5652K}, we evaluate SMBH masses by assuming that a fraction of the central stellar mass collapses into the SMBH. Finally, we compare the simulated SMBH masses to observed scaling relations to assess whether MOND-modelled ETGs can host SMBHs consistent with empirical data. This work builds on the foundation of \citet{2022MNRAS.516.1081E} and aims to connect the dynamics of MOND-modelled ETGs with SMBH formation, providing new insights into galaxy evolution in alternative gravity frameworks.

\section{Model analysis method}
\label{sec:models}
In this study, we use the model galaxies formed with an initial post-Big-Bang gas cloud radius (\(R_{\text{initial}}\)) of 500 kpc as presented in \citet{2022MNRAS.516.1081E}. Here, we specifically focus on analysing \(\Phi_{\text{central}}\), gas inflow rates, and estimating the potential mass of a SMBH that could form within the galaxy. For these analyses, we employ the model "e38," which is run at a higher temporal resolution compared to other models. This decision is motivated by the need to capture detailed gas dynamics and potential evolution during the critical phases of galaxy formation. However, due to computational limitations, we have restricted our analysis to this single high temporal resolution model. For estimating the potential mass reservoir available for SMBH formation, we use the gas mass evolution data from our main model (e38), which employs an initial cloud radius \( R_{\text{initial}} = 500 \, \text{kpc} \). Although we mention multiple models with this initial radius (e35, e36, e37, e38 and e39 from Table 1 of \citealt{2022MNRAS.516.1081E}), our SMBH mass estimate is based specifically on the e38 model. The time reported here is relative to the start of the simulation. The simulation box has a side length of 1 Mpc, providing a sufficient volume to follow the collapse and evolution of the isolated galaxy model. Throughout the analysis, we define the 'central region' based on the centre of mass of the stellar particles within a 500 kpc radius. This approach ensures that the centre is dynamically tracked throughout the simulation, independent of any fixed box coordinates.

The simulations are further constrained by a maximum spatial resolution of 0.24 kpc (which is the physical size of an individual grid cell at the highest level of refinement. However, we note that accurately resolving potential gradients typically requires several grid cells. Thus, the effective resolution of our simulation is somewhat larger than the cell size, approximately $\approx$1 kpc in the densest regions.), determined by the adaptive mesh refinement (AMR) capabilities of the PoR code. Improving the spatial resolution would provide more accurate results, but this is currently infeasible due to the significant computational cost associated with higher resolution MOND simulations.While the PoR code in QUMOND mode -- the quasi-linear formulation of MOND \citep{2010MNRAS.403..886M} in which the modified gravitational field is derived via a two-step Poisson solver indeed requires at most twice the computational time compared to Newtonian runs, the primary limitation for increasing refinement levels in our simulation was the available memory capacity. Higher refinement would substantially increase memory requirements due to the number of cells and particles, and future work will aim to address this with improved computational resources. Despite these limitations, the model allows us to explore key aspects of SMBH formation and galaxy evolution within the framework of MOND. A detailed description of the model formation process, including the initial conditions and simulation setup, can be found in \citet{2022MNRAS.516.1081E}.

\subsection{The gravitational potential at the centre of the model relic}
The gravitational potential at the centre of the model, \(\Phi_{\text{central}}\), is calculated in the simulation by solving the MOND-modified Poisson equation. In MOND, the gravitational potential \(\Phi\) is governed by the equation:

\begin{equation}
\Delta \Phi = 4 \pi G \rho_b + \nabla \cdot \left[\tilde{\nu} \left(\frac{|\nabla \Phi|}{a_0}\right) \nabla \Phi\right],
\label{eq:mond_potential}
\end{equation}
where \(\Delta \Phi\) is the Laplacian of the gravitational potential, \(\rho_b\) is the baryonic matter density, \(G\) is the gravitational constant, \(a_0\) is the MOND acceleration constant, and \(\tilde{\nu}(y)\) is the MOND transition function, with \(y = |\nabla \Phi| / a_0\). This equation represents the formulation of the Poisson equation in the QUMOND framework, as implemented in the PoR code. This equation modifies Newtonian gravity at low accelerations (\(a < a_0\)) by introducing non-linearity in the relationship between mass and potential. The PoR code numerically solves this equation to compute the gravitational potential at various points within the galaxy.

The calculation begins with the baryonic density distribution (\(\rho_b\)), which represents the distribution of gas and stars in the galaxy. Using this density, the Newtonian Poisson equation is first solved to calculate the Newtonian potential \(\phi(r)\):

\begin{equation}
\Delta \phi = 4 \pi G \rho_b.
\label{eq:newtonian_potential}
\end{equation}
From the Newtonian potential, the code computes the phantom dark matter density (\(\rho_{\text{ph}}\)) that arises due to the MOND modification. This density is given by:

\begin{equation}
\rho_{\text{ph}} = \frac{1}{4\pi G} \nabla \cdot \left[\tilde{\nu} \left(\frac{|\nabla \phi|}{a_0}\right) \nabla \phi \right].
\label{eq:phantom_dm_density}
\end{equation}
The total gravitational potential \(\Phi\) is then recalculated by solving the Poisson equation, which incorporates both \(\rho_b\) and \(\rho_{\text{ph}}\):

\begin{equation}
\Delta \Phi = 4 \pi G (\rho_b + \rho_{\text{ph}}).
\label{eq:total_potential}
\end{equation}
\(\Phi_{\text{central}}\) is extracted from the simulation output at specific radii, such as 1 kpc, 0.8 kpc, and 0.6 kpc, at each time step (The simulation outputs are saved at intervals of 10 Myr, allowing us to follow the evolution of inflow rates and dynamical changes over time. For the period of rapid collapse ($\approx$3–5 Gyr), this corresponds to a temporal resolution of 100 snapshots per Gyr). This value represents the depth of the gravitational potential well at the galaxy's centre and evolves as the galaxy forms and stabilizes.

Physically, \(\Phi_{\text{central}}\) reflects the gravitational strength in the galaxy's core. A deeper potential well corresponds to a denser and more massive central region. The evolution of \(\Phi_{\text{central}}\) over time provides insight into the baryonic collapse and the formation of a dense core, processes that are critical for driving gas inflows and forming a SMBH. Additionally, the differences in \(\Phi_{\text{central}}\) at various radii, such as between 1 kpc and 0.6 kpc, reveal the steepness of the potential gradient, which influences the dynamics of gas inflows and central star formation.

By solving the MOND-modified Poisson equation and extracting gravitational potential values at central radii, the simulation provides a detailed measure of \(\Phi_{\text{central}}\), which is fundamental to understanding galaxy evolution and the conditions for SMBH formation in MOND.

\subsection{Gas infall at the centre of the model relic}
The gas inflow rates at different radii (\(r = 1 \, \text{kpc}, 0.8 \, \text{kpc}, 0.6 \, \text{kpc}\)) were computed using data from the simulation. The inflow rate (\(\dot{M}_{\rm gas}\)) was calculated as the rate of change of gas mass (\(M_{\rm gas}\)) within a given radius over a discrete time interval. For each radius, the inflow rate is given by:

\begin{equation}
\dot{M}_{\rm gas}(r, t) = \frac{\Delta M_{\rm gas}(r, t)}{\Delta t},
\end{equation}
where \(\dot{M}_{\rm gas}(r, t)\) is the inflow rate at radius \(r\) and time \(t\), \(\Delta M_{\rm gas}(r, t) = M_{\rm gas}(r, t + \Delta t) - M_{\rm gas}(r, t)\) is the change in gas mass within radius \(r\) between consecutive time snapshots, \(\Delta t\) is the time interval between these snapshots. The gas mass is calculated by summing the contributions from individual cells within the specified radius. For each cell, the gas mass is obtained by multiplying the cell volume by its gas density. Cells whose centres lie within the selected radius are included in full. Since the AMR grid structure results in varying cell resolutions, this calculation automatically accounts for local resolution differences. We note that we do not perform partial cell volume calculations for boundary cells; cells are either fully included or excluded based on their centre position.

The simulation data provides \(M_{\rm gas}(r, t)\) for discrete time points. Using the differences between consecutive snapshots, the inflow rate is computed for the midpoints of these time intervals to align with the calculated rates.

\subsection{Black hole mass estimation}
This analysis estimates the black hole mass (\(M_{\rm BH}\)) as a fraction of the central stellar mass and computes the central velocity dispersion (\(\sigma\)) for galaxies within the simulation. The results are compared to the observed \(M_{\rm BH} - \sigma\) scaling relation from \citet{2013ApJ...764..184M}.

The central stellar mass (\(M_{\rm central}\)) is calculated by summing the masses of all stellar particles formed from the gas located within a defined radius (\(r_{\rm central} = 1 \, \text{kpc}\)) of the galaxy centre. The distance of each particle from the centre is computed as:

\begin{equation}
r = \sqrt{x^2 + y^2 + z^2},
\label{eq:distance}
\end{equation}
where \(x, y, z\) are the particle coordinates. Particles with \(r \leq r_{\rm central}\) are selected, and their masses are summed to obtain the central mass:

\begin{equation}
M_{\rm central} = \sum_{i}^{N} m_i,
\label{eq:central_mass}
\end{equation}
where \(m_i\) is the mass of the \(i\)-th particle in the central region.

The black hole mass is estimated as a fraction of the central stellar mass. For a given fraction \(f\), the black hole mass is given by:

\begin{equation}
M_{\rm BH} = f \times M_{\rm central}.
\label{eq:bh_mass}
\end{equation}

Fractions ranging from \(f = 0.1\) to \(f = 1.0\) are used to explore the contribution of the central stellar mass to the SMBH. These fractions are realistic \citep{2020MNRAS.498.5652K} as the stellar initial mass function (IMF) depends on the metallicity and strongly on the star-forming gas and is top-heavy under the relevant physical conditions such that the majority of the cluster's mass ends up being in stellar mass black holes (\citealt{2012MNRAS.422.2246M, 2018A&A...620A..39J, 2020MNRAS.498.5652K, 2024arXiv241007311K}).

The 3D central velocity dispersion (\(\sigma_{\rm total}\)) is calculated using the particle velocities in the central region. For each velocity component (\(v_x, v_y, v_z\)), the dispersion is computed as:

\begin{equation}
\sigma_x = \sqrt{\frac{1}{N} \sum_{i=1}^{N} (v_{x, i} - \bar{v}_x)^2},
\label{eq:velocity_dispersion_component}
\end{equation}
where \(N\) is the number of particles, \(v_{x, i}\) is the velocity of the \(i\)-th particle, and \(\bar{v}_x\) is the mean velocity in the \(x\)-direction. Similar expressions are used for \(\sigma_y\) and \(\sigma_z\). The total 3D central velocity dispersion is,

\begin{equation}
\sigma = \sqrt{\sigma_x^2 + \sigma_y^2 + \sigma_z^2}.
\label{eq:total_velocity_dispersion}
\end{equation}

The estimated \(M_{\rm BH}\) and the computed velocity dispersion values are compared to the observed \(M_{\rm BH} - \sigma\) relation from \citet{2013ApJ...764..184M}:

\begin{equation}
\log_{10}(M_{\rm BH}) = 8.39 + 5.20 \cdot \log_{10} \left(\frac{\sigma}{200}\right),
\label{eq:scaling_relation}
\end{equation}
where \(\sigma\) is in \(\text{km/s}\). This comparison provides insight into whether the simulated galaxies reproduce the observed scaling relation between black hole mass and central velocity dispersion, thereby validating the model's predictions for SMBH formation.

\section{Results}
\label{sec:results}

\subsection{Is the potential well deep enough to accommodate a SMBH?}
\begin{figure*}

\includegraphics[scale=0.6]{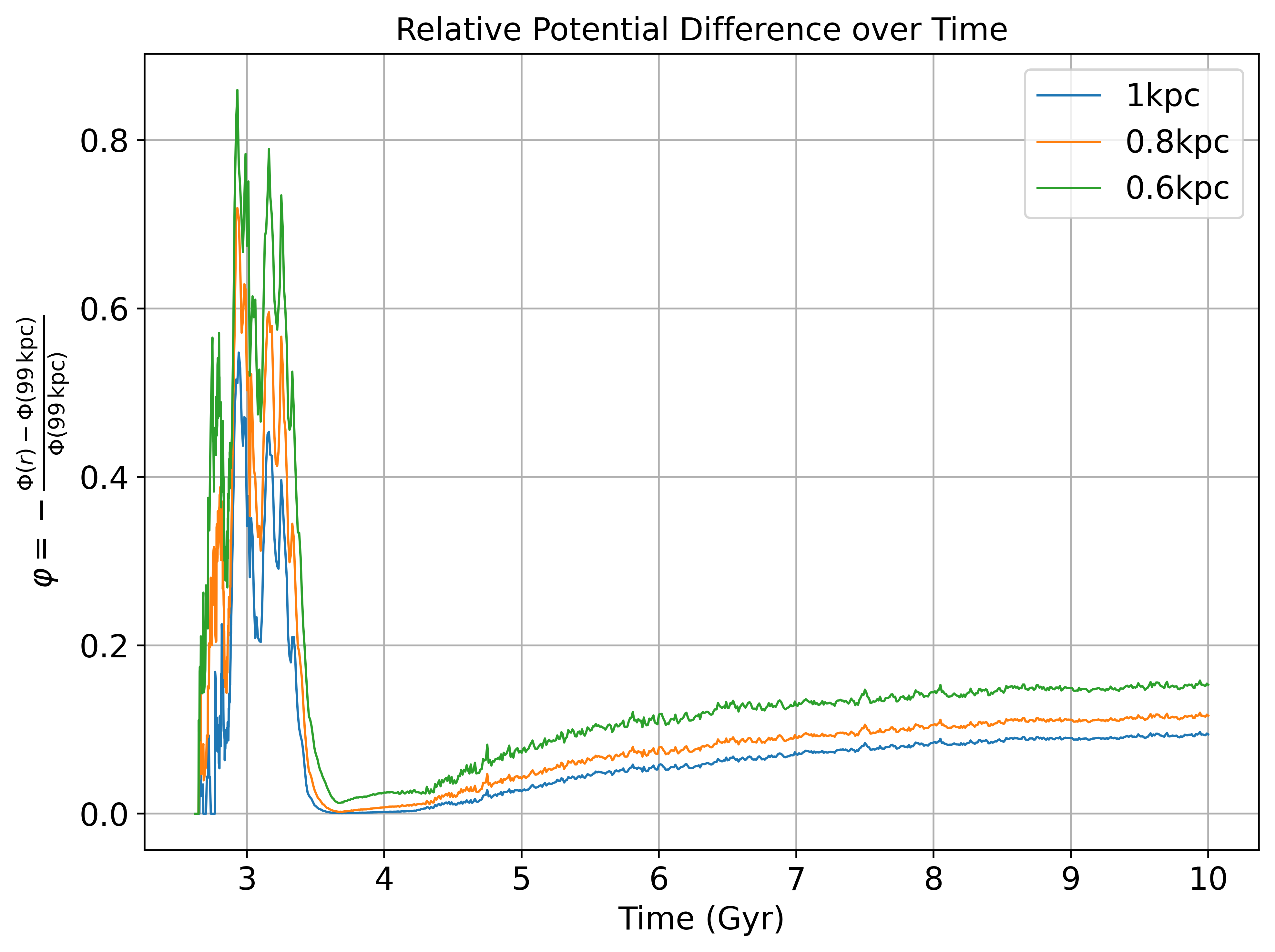}
\\
    \caption{Time evolution of the relative gravitational potential difference between the central region (at radii of 1 kpc, 0.8 kpc, and 0.6 kpc) and a reference point at 99 kpc. The potential deepens rapidly during the collapse phase ($\approx$ 3–5 Gyr), stabilizing thereafter. The use of potential difference avoids dependence on absolute values and reflects the physical depth of the potential well relative to the galaxy outskirts.}

    \label{fig:potential}
\end{figure*}

We analyse the evolution of the gravitational potential difference at the centre of our model ETG (e38), which forms a total stellar mass of \(0.6 \times 10^{11} \, M_\odot\). Figure \ref{fig:potential} shows the evolution of the relative gravitational potential difference relative to the potential at a radius of 99 kpc between three central radii (1 kpc, 0.8 kpc, and 0.6 kpc) over 10 Gyr.

The reference radius at 99 kpc corresponds to the outskirts of our simulated system, providing a consistent frame of reference for measuring the depth of the central potential well. While an ideal reference would be at infinity, the finite simulation volume necessitates this practical choice. The potential difference highlights the deepening of the gravitational well relative to the galaxy outskirts.

During the initial 4–5 Gyr, the relative potential difference increases sharply, coinciding with the collapse of baryonic gas and the formation of the stellar population. This steep rise reflects the rapid deepening of the central gravitational well, driven by gas condensation and intense star formation activity. This phase is critical as it establishes the gravitational environment that facilitates further mass inflow toward the galactic center.

Beyond 5 Gyr, the potential difference stabilizes, indicating that the system transitions to a more dynamically settled state. Notably, the deepest potential difference is seen at smaller radii (0.6 kpc), emphasizing the concentration of baryonic mass toward the centre and suggesting that any gas reaching these inner regions could contribute to black hole growth.

\begin{figure*}
\includegraphics[scale=0.6]{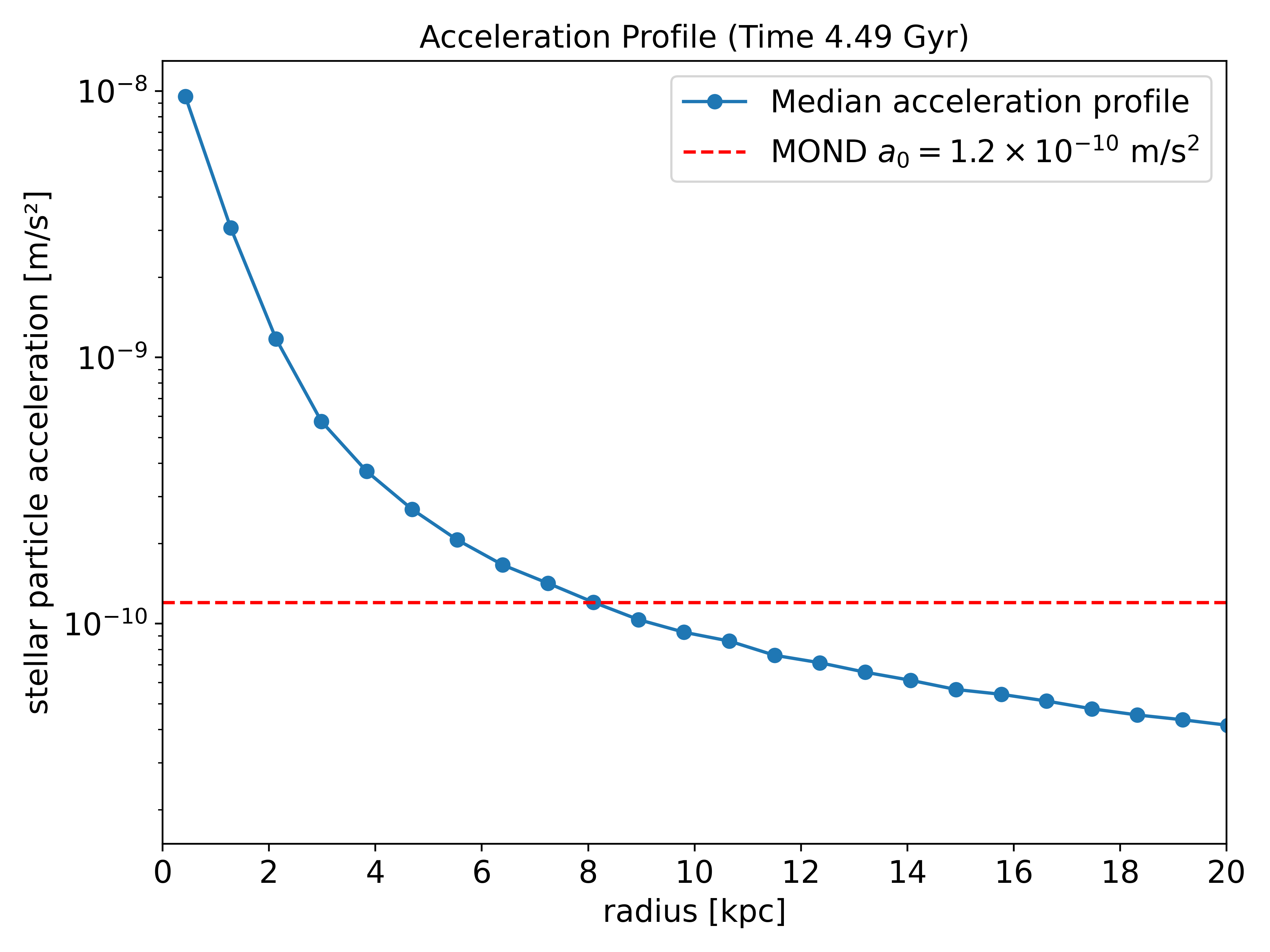}
\\
    \caption{Radial profile of the median stellar particle acceleration at \(t = 4.49 \, \text{Gyr}\). The red dashed line indicates the MOND acceleration scale \(a_0 = 1.2 \times 10^{-10} \, \text{m/s}^2\) $\approx$ 3.8 pc/Myr$^{2}$. Accelerations in the inner $\approx$3–4 kpc exceed \(a_0\), indicating that the central region is in the Newtonian regime. MOND effects emerge at larger radii ($\approx$8 kpc), influencing the global gravitational collapse.}
    \label{fig:acceleration}
\end{figure*}

To assess the role of MOND dynamics in the evolution of the central potential, we computed the acceleration profile at \(t = 4.49 \, \text{Gyr}\), corresponding to the peak of baryonic collapse (Figure \ref{fig:acceleration}). The accelerations within the inner $\approx$ 3–4 kpc exceed the MOND acceleration scale \(a_0\), indicating that the dynamics in the central potential well remain Newtonian due to the high baryonic density. MOND effects become significant at larger radii ($\approx$ 8 kpc and beyond), potentially aiding the gravitational collapse and mass transport toward the centre.

A comparable observational example is the compact ETG NGC 1277 in the Perseus Cluster, with a total stellar mass of \(1.2 \times 10^{11} \, M_\odot\) and an effective radius of approximately 1.2 kpc. This galaxy hosts a SMBH with a mass of \(\approx 1.7 \times 10^{10} \, M_\odot\), accounting for nearly 14\% of its total stellar mass — far above typical scaling relations. The gravitational potential difference in NGC 1277 is similarly deep, driven by its compact stellar core and massive black hole \citep{2014ApJ...780L..20T, 2012Natur.491..729V, 2017MNRAS.468.4216Y, 2017MNRAS.467.1929F}. This is comparable to the potential difference we observe in our model after $\approx$ 5 Gyr, supporting the notion that such compact systems naturally develop deep gravitational wells early in their evolution.

This comparison highlights that the gravitational environment in our simulated relic galaxy aligns well with observed properties of real compact ETGs like NGC 1277. The deep central potential in both cases underscores the role of dense baryonic cores in establishing conditions favourable for SMBH formation and growth. The evolution of the potential difference in our model further supports the idea that compact ETGs can host over-massive SMBHs as a consequence of their early, rapid formation and subsequent stabilization.

\subsection{Is there sufficient gas infall to fuel the seeds?}
\begin{figure*}

\includegraphics[scale=0.6]{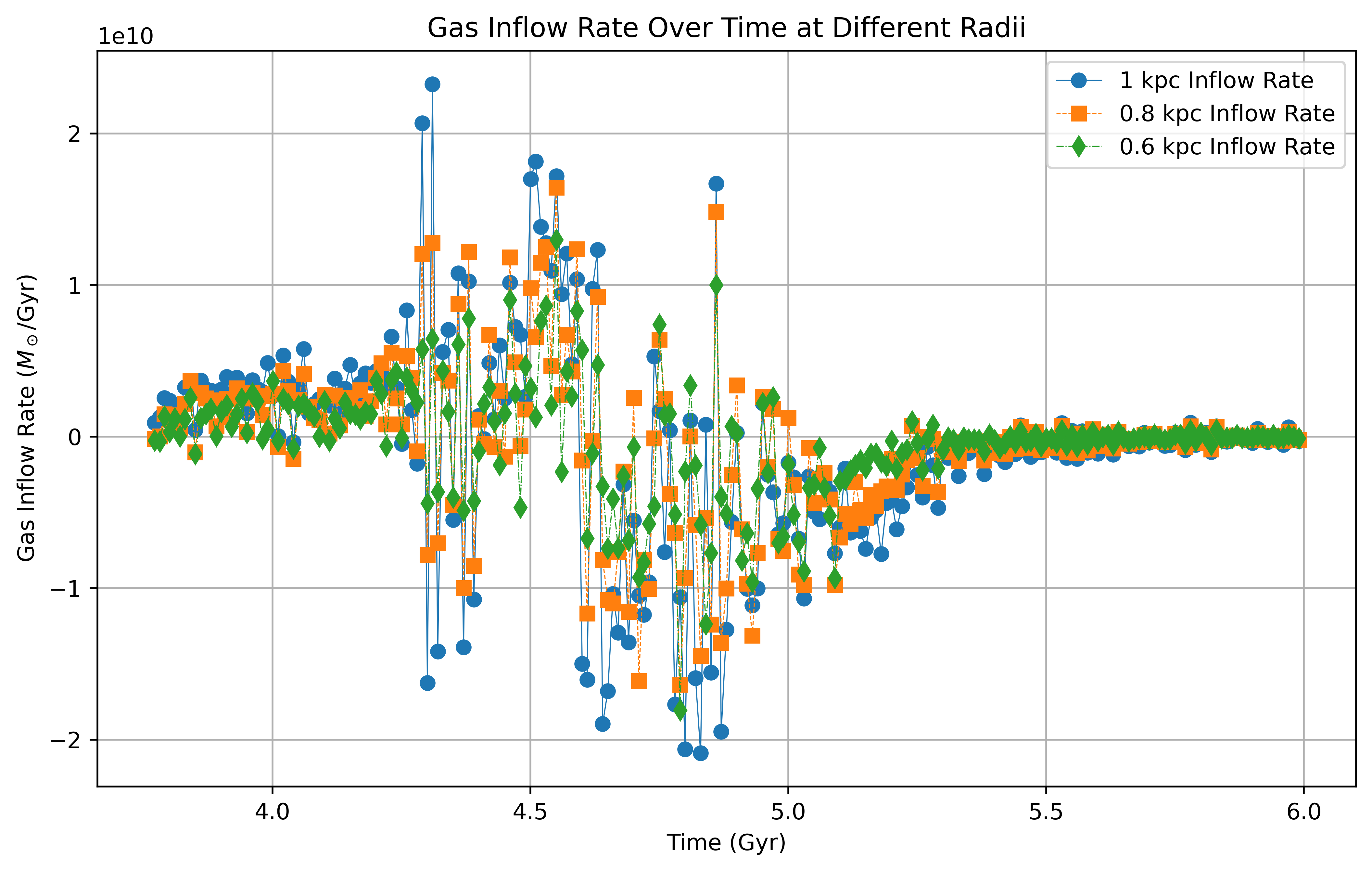}
\\
    \caption{Gas inflow rate over time at different radii (1 kpc, 0.8 kpc, and 0.6 kpc). The plot illustrates fluctuations in the inflow rate of gas (in \( M_\odot/\text{Gyr} \)) during the time range from 4 Gyr to 6 Gyr. Significant variability is observed, particularly in the interval between 4.0 and 5.0 Gyr, with notable peaks and troughs, indicative of dynamic processes affecting gas inflow at these radii.}

    \label{fig:gasinflow}
\end{figure*}
The gas inflow rates at the same radii (1 kpc, 0.8 kpc, and 0.6 kpc) are illustrated in Figure \ref{fig:gasinflow} for the critical epoch between 4 and 6 Gyr, corresponding to the period of most rapid potential evolution. At a radius of 1 kpc, gas inflow peaks at rates exceeding \(2 \times 10^{10} \, M_\odot/\text{Gyr}\), while smaller radii exhibit similarly high but slightly delayed peaks. These inflows correspond to periods of rapid cooling and contraction of gas toward the center. The sustained high inflow rate over about half a Gyr suggests that sufficient mass can be channelled into the inner regions to seed and grow a central SMBH. Notably, around \(t \approx 4.7 \, \text{Gyr}\), the central regions experience an outflow phase, as shown in Figure \ref{fig:gasinflow}, driven by feedback mechanisms inherent to the model (\citealt{2022MNRAS.516.1081E}), leading to gas heating and temporary disruption of inflows. After 5 Gyr, outflow rates decline and stabilize, consistent with the exhaustion of central gas reservoirs. This reduction implies that star formation and black hole accretion activity slow significantly, transitioning the system into a quiescent phase.

These findings underscore the importance of the early epochs (4–5 Gyr, i.e., 0.5-1.5 Gyr after start of collapse, see Figure \ref{fig:potential}) as a critical window for black hole growth, during which high inflow rates enable the accretion of substantial mass into the core. We note that during the time $\approx$ 3.7 - 4.7 Gyr there is a slowly rising infall rate from 0.1  \(\times 10^{10}\ M_\odot/\text{Gyr} \) to 1.1 \(\times 10^{10} \ M_\odot/\text{Gyr} \). The inflow rates exhibit substantial variability after about 4.5 Gyr with fluctuations between strong inflow and outflow on timescales comparable to our snapshot interval driving supernova activity. Given the limited spatial and temporal resolution of our simulation, and the inherent variability in the gas inflow rates, our conclusions regarding the viability of SMBH seed formation remain qualitative. Higher resolution simulations would be required to fully resolve the detailed inflow behaviour at sub-kpc scales relevant for SMBH feeding.

The trends observed in our model align closely with theoretical predictions and observations of ETGs in the literature. For instance, \citet{2020MNRAS.498.5652K} proposed that SMBHs form through the coalescence of stellar-mass black holes in dense star clusters. These clusters, formed during the collapse of gas-rich systems, rely on significant gas inflows to maintain their density and fuel black hole mergers. The inflow rates in our simulation, peaking above \(2 \times 10^{10} \, M_\odot/\text{Gyr}\) at 1 kpc, provide the necessary conditions for this process, allowing central gas densities to remain high enough to support the formation of the massive clusters. Furthermore, the slight delay in peak inflows at smaller radii (e.g., 0.6 kpc) highlights the gradual inward transport of gas, a feature consistent with the cooling flows observed in present day dense early environments.

Observational studies of high-redshift quasars (e.g., \citealt{2013ApJ...779...24V}) have revealed black holes with masses exceeding \(10^9 \, M_\odot\) by \(z \approx 6\), implying efficient accretion mechanisms in the first billion years. This comparison reinforces the notion that the early, gas-rich environment in our simulated ETG is consistent with the conditions required for forming and growing central SMBHs observed in compact ETGs and high-redshift quasars.

\subsection{Would the $M_{\rm BH} - \sigma$ scaling relation emerge?}
\begin{figure*}

\includegraphics[scale=0.49]{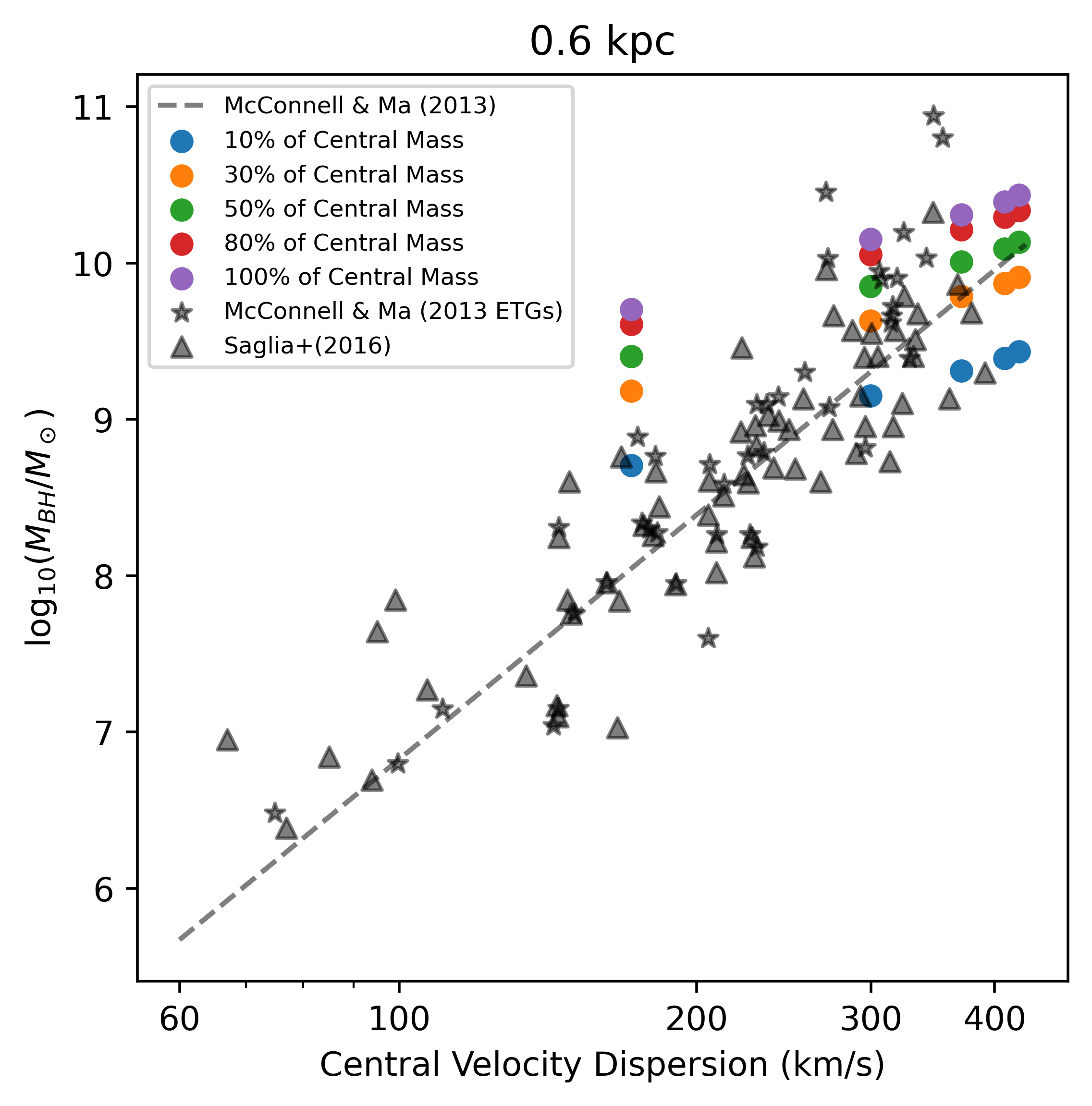}
\includegraphics[scale=0.49]{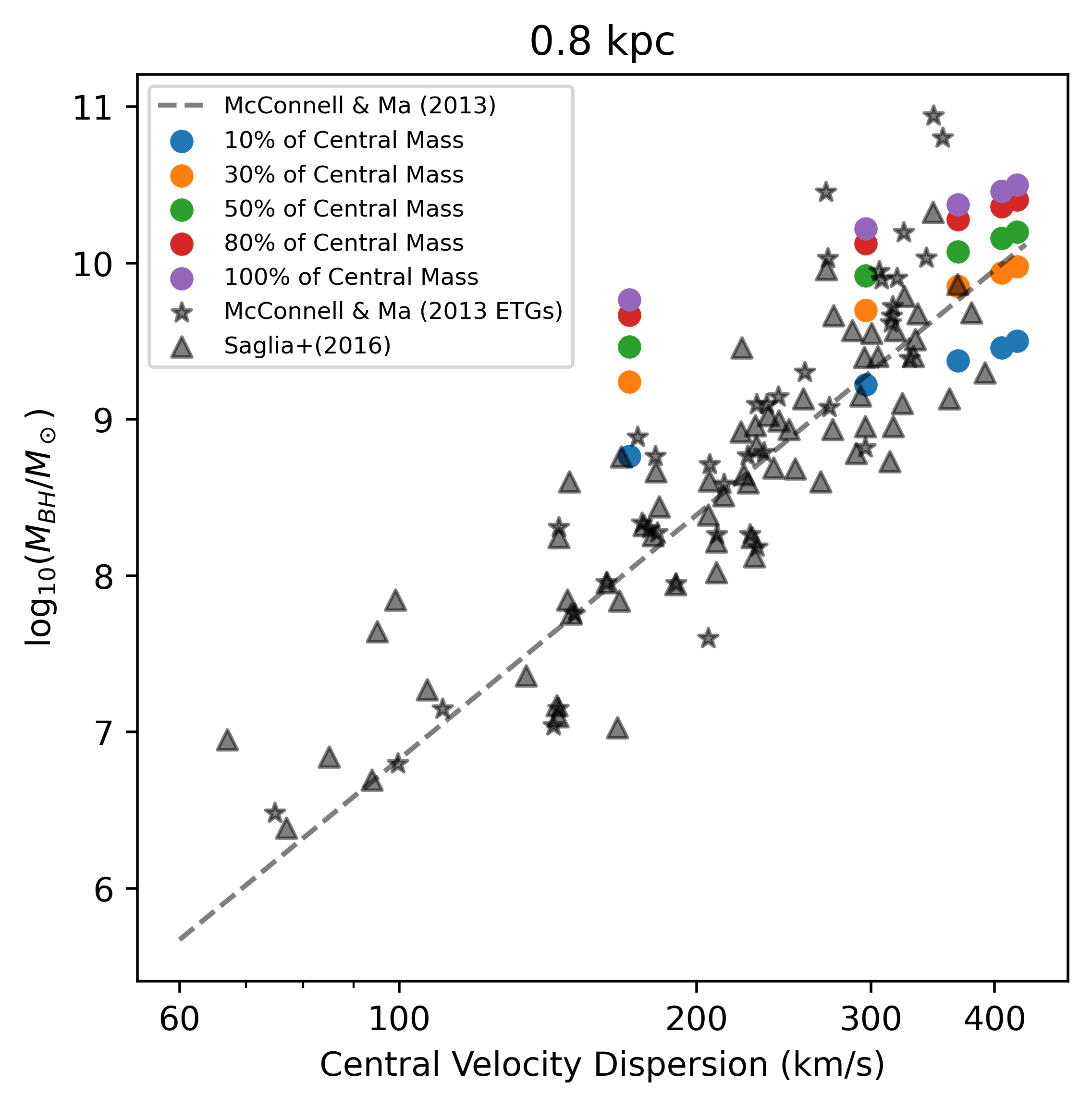}
\includegraphics[scale=0.49]{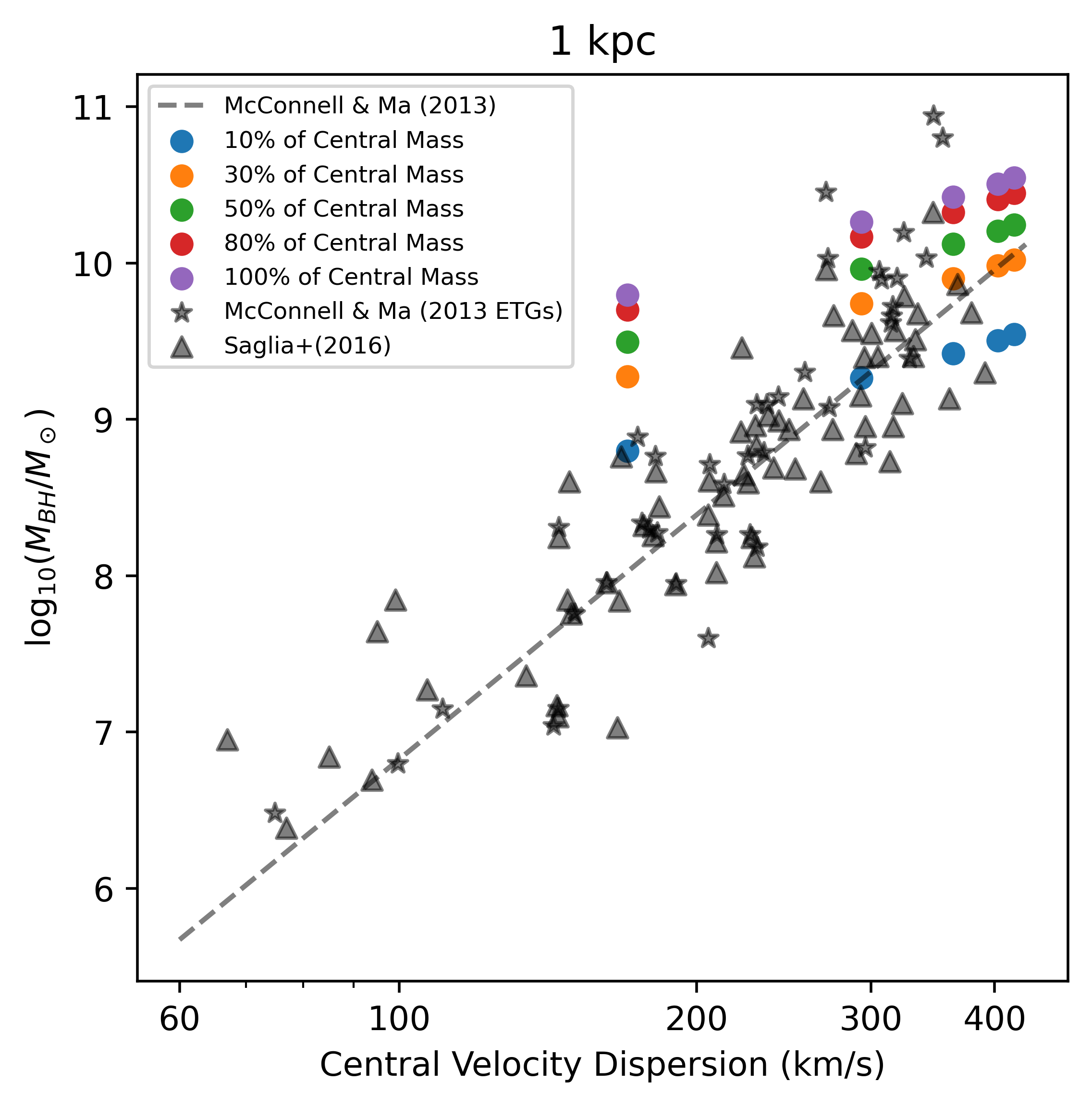}
\\
\caption{The calculated black hole masses for the model relics (e35, e36, e37, e38 and e39) from \citet{2022MNRAS.516.1081E} , based on 10$\%$, 30$\%$, 50$\%$, 80$\%$, and 100$\%$ of the central stellar mass, are plotted alongside the observed \(M_{\rm BH}-\sigma\) relation from \citet{2013ApJ...764..184M}. The plotted black stars and black triangles are the ETGs from \citet{2013ApJ...764..184M} and \citet{2016ApJ...818...47S} respectively.}

    \label{fig:3dveldisp}
\end{figure*}

The relationship between \(M_{\rm BH}\) and \(\sigma\) is explored in the context of our simulation by assuming that a certain fraction of the central stellar mass within 1 kpc, 0.8 kpc, and 0.6 kpc contributes to black hole formation. This is based on the model of \citet{2020MNRAS.498.5652K} in which the most massive star clusters form in the innermost regions of the forming spheroidal galaxy, each 10 Myr. The calculated black hole masses, based on 10 $\%$, 30 $\%$, 50 $\%$, 80 $\%$, and 100 $\%$ of the central stellar mass, are plotted alongside the observed \(M_{\rm BH}-\sigma\) relation from \citet{2013ApJ...764..184M}. The results shown in Figure \ref{fig:3dveldisp} include the models  e35, e36, e37, e38 and e39 from \citet{2022MNRAS.516.1081E} as well as the observed ETGs from \citet{2013ApJ...764..184M} and \citet{2016ApJ...818...47S}.

Black hole masses derived from lower contributions of the central stellar mass (10–50$\%$) and at larger central velocity dispersion exhibit good agreement with the observed \(M_{\rm BH}-\sigma\) relation across all radii. This suggests that if a significant fraction of the central stellar mass contributes to the black hole's formation, then the resulting masses are consistent with those observed in real ETGs. This indicates that a minimum threshold of stellar mass contribution is necessary to match observed SMBH properties in excellent agreement with the analytical results of \citet{2020MNRAS.498.5652K}. We acknowledge that the spatial scales considered in our analysis (0.6–1 kpc) are significantly larger than the physical scales relevant for SMBH accretion, which are typically in the parsec range. Therefore, our estimates represent upper limits to the mass reservoir potentially available for SMBH formation, and the actual mass feeding the black hole would likely be orders of magnitude smaller.

The analysis highlights the capability of MOND dynamics to support the formation of SMBHs in ETGs formed via monolithic collapse. The rapid deepening of the central gravitational potential during the first 0.5 Gyr after begin of the collapse, coupled with high gas inflow rates, creates a conducive environment for black hole seeding and growth.

\section{Conclusion}
\label{sec:conclusion}

In this study, we investigated the evolution of ETGs within a MOND framework, focusing on the central gravitational potential, gas inflow rates, and the relationship between black hole mass and central velocity dispersion. Using simulations, we analysed the gravitational and dynamical properties of a model galaxy with a total stellar mass of \(0.6 \times 10^{11} \, M_\odot\), providing insights into the mechanisms governing the formation and growth of SMBHs in compact systems.

The evolution of the central gravitational potential reveals a sharp decline during the first 0.5 Gyr after the start of the collapse, corresponding to the rapid collapse of baryonic matter and the formation of the stellar population. This phase is marked by the deepening of the central potential well. The deeper gravitational potential at smaller radii (e.g., 0.6 kpc) underscores the concentration of baryonic matter in the central regions, establishing the conditions necessary for sustained gas inflows and SMBH formation.

Gas inflow rates during the critical epoch between 0.5 and 1.5 Gyr after the collapse starts exhibit peaks exceeding \(2 \times 10^{10} \, M_\odot/\text{Gyr}\) at 1 kpc, with similarly high but slightly delayed peaks at 0.8 kpc and 0.6 kpc. The subsequent outflow rates and their decline rates after 4.7 Gyr reflects stellar activity and the depletion of gas reservoir, transitioning the system into a quiescent phase. Variations in the adopted feedback prescriptions could significantly influence the inflow dynamics and star formation rates, potentially altering the efficiency of central mass accumulation. Exploring these effects will be the subject of future work. This early period of high inflow rates is crucial for fuelling the growth of the central SMBH and forming a dense stellar core, consistent with the analytical results by \citet{2020MNRAS.498.5652K}.

Finally, the relationship between \(M_{\rm BH}\) and \(\sigma\) was examined by assuming various fractions (10$\%$, 30$\%$, 50$\%$, 80$\%$, and 100$\%$) of the central stellar mass contribute to the black hole. Black hole masses derived from this align closely with the observed \(M_{\rm BH}-\sigma\) relation, particularly at moderate to high velocity dispersions (\(\sigma > 200 \, \text{km/s}\)). The results suggest that the combination of a deep gravitational potential, high early gas inflow rates, and significant central mass contributions is essential for forming SMBHs consistent with those observed in real galaxies. A more accurate estimate of the SMBH formation potential would require resolving sub-parsec scales and incorporating detailed accretion physics. While our current resolution limits such analysis, the observed collapse and mass assembly provide an upper-limit framework for assessing early SMBH formation in MOND.

Overall, our findings demonstrate that ETGs formed within the MOND framework have the potential to naturally reproduce many of the observed properties of ETGs, including their dense cores, quiescent phases, and the \(M_{\rm BH}-\sigma\) relation. These results underscore the importance of early gas inflows and gravitational potential evolution in shaping the growth of SMBHs and the structural properties of ETGs, offering a viable pathway for understanding their formation and evolution.

However, we recognize certain limitations in this work. The radii typically around 0.1 kpc or smaller, are not fully resolved in our simulations due to computational limitations. As a result, the work presented here provides only a basic estimate of the conditions for SMBH formation and growth in MOND. This is a simplified approach that does not account for complex processes such as detailed feedback mechanisms, the actual coalescence of the cluster of stellar remnants to a SMBH seed, relativistic effects, or accretion dynamics at smaller scales. Nonetheless, this work offers a foundational framework for future MOND cosmological simulations, highlighting the importance of early gas inflows and gravitational potential evolution in shaping compact ETGs and their central SMBHs. Future efforts to incorporate finer resolutions and additional physics into MOND-based simulations will be critical for achieving a more comprehensive understanding of these processes.

\begin{acknowledgement}
The author R.E would like to thank KFPP for their support. We thank the DAAD Bonn - Eastern European Exchange programme at the University of Bonn and Prague for support.\end{acknowledgement}

\printendnotes
\printbibliography
\end{document}